\documentstyle[aps,epsf]{revtex}
\begin{document}
\begin{twocolumn}
\title{Comment on ``One-Step Synthesis of Multiatom 
Greenberger-Horne-Zeilinger States''}
\author{G.S. Agarwal$^1$, R.R. Puri$^2$, and R.P. Singh$^1$}
\address{$^1$Physical Research Laboratory, Ahmedabad-380009,
$^2$Bhabha Atomic Research Center, Mumbai-400085}
\date{\today}
\maketitle
In a recent Letter, Zheng \cite{1.zeng} shows a very interesting
method of generating
multiatom 
Greenberger-Horne-Zeilinger (GHZ) states. He considers a model system
consisting 
of a set of N two-level atoms interacting with the vacuum of the single mode 
cavity. The cavity is assumed to be sufficiently detuned so that the cavity 
does not cause real transitions between the excited and the ground state of 
the atom. He acknowledges  that the effective Hamiltonian derived by him is 
the same as the one derived in our paper (APS) \cite{2.Agarwal}. He then
studies the time 
evolution of the atomic system. He specifically demonstrates the generation 
of the GHZ states for $\eta t= \pi/2$. His most general result is given by
the 
Eq. (49) of which his other key results (15) and (30) are special cases, 
obtained by setting N=3 and 4 in Eq. (49). Our earlier paper
\cite{2.Agarwal} gives the 
time evolution of the system for a very general initial condition using the 
same Hamiltonian. One of our key results is given by Eq. (22) of APS. 
This particular result of APS shows that at time $\eta t=\pi/2$ the state
of the 
system is an atomic cat state if the initial state of the system is chosen 
as an atomic coherent state $|\theta,\phi\rangle$. The purpose of this
comment
is to 
point out the relation between the works of Zheng and APS and to show the
important connection between the atomic cat state and the GHZ states. 
Zheng's key result Eq. (49) is found to be a special case of our general
result
Eq. (22) for the 
choice of the parameters $\theta=\pi/2$ and $\phi=-\pi/2$. We further
discuss the 
detection of the atomic cat state by using the Ramsey 
spectroscopy. 

The fact that the results of references \cite{1.zeng} and
\cite{2.Agarwal} are identical should
not 
come as a surprise and is only expected if one considers the same interaction 
Hamiltonian, same initial state and the same value of the interaction time. 
Needless to say that our paper goes much beyond what is shown in the work 
of Zheng  as we choose a variety of initial states and interaction times 
- see, for example, the more general entangled states described by 
Eqs. (20, 21) of APS. Besides we discuss the quantum character of such states 
by examining 
the quasidistributions for these states [Ref. \cite{3.gsa} which
also
discusses 
the Wigner function and the decoherence  of such entangled states]. 
The interference character shown in the Fig. 3 of APS is a reflection
of 
the entangled character of the state. It might be added that dispersive 
interaction of the type discussed by APS is quite natural in the context of 
Bose condensates where one has begun to produce entangled states.
Let us now demonstrate explicitly what we stated above. The effective
Hamiltonian for the problem is $\eta S^+S^-$ where $S^\pm$ are the
collective operators for a system of N two-level atoms. The initial state
considered by Zheng is $|\psi(0)\rangle=\prod_j
\frac{1}{\surd 2}(|g_{j}>+i|e_{j}\rangle)$
and he shows that the state at $\eta t=\pi/2$ is a multiparticle GHZ state
(Eq. (49) of Ref. \cite{1.zeng})
\begin{eqnarray}
|\psi(t)\rangle & = & \frac{e^{i\pi/4}}{\surd 2}\{\prod_{j}
\frac{1}{\surd 2}(|g_{j}\rangle +(-i)^{N}|e_{j}\rangle) \nonumber \\
&   &\mbox{}-i\prod_{j}\frac{1}{\surd 2}(|g_{j}\rangle-(-i)^{N}
|e_{j}\rangle)\}
\end{eqnarray}
The APS result for an initial state that is atomic coherent state
$|\theta,\phi\rangle$ is
\begin{eqnarray}
|\psi\rangle_{A} & \equiv & \frac{e^{-iN\pi/2}}{\surd
2}[e^{i\pi/4}|\theta,\phi-\frac{\pi(N-1)}{2}\rangle\nonumber \\
&   &\mbox{}+e^{-i\pi/4}|\theta,\phi-\frac{\pi(N-3)}{2}\rangle]
\end{eqnarray}
For the fully symmetric case considered by APS the atomic coherent state
is an uncorrelated state $|\theta,\phi\rangle =
e^{iN\phi}\prod_{j}\left(\cos\frac{\theta}{2}|g_{j}\rangle +
e^{-i\phi}\sin\frac{\theta}{2}|e_{j}\rangle\right).$ By straightforward
algebra one can prove that $|\psi\rangle_{A} =
e^{-iN\pi/2}|\psi\rangle.$ Note that $|\psi(0)\rangle_{A} =
e^{-iN\pi/2}|\psi(0)\rangle$.

Consider next the detection of states like (1) or (2). For this purpose we
can use a Ramsey set up. In the first Ramsey zone (just
before atoms
enter the cavity) the atoms are prepared in the state
$|\theta,\phi\rangle$. In the second Ramsey zone just after the cavity the
cat state (2) gets projected onto $\langle \alpha,\beta|\psi\rangle$,
where $\alpha$ and $\beta$ refer to the parameters of the external field
in the second Ramsey zone. After the second Ramsey zone the probability of
detecting all atoms in the ground state would yield the entangled nature
of the state (2). In particular for $\theta =\pi/2,\phi =-\pi/2$ and for
$\alpha = \pi/2$, N= odd, Eq. (2) yields an interference pattern
that is quite different from the corresponding results in the absence of
the cavity and for the incoherent mixture. Note
that one can produce a variety of interference patterns for the state
(2) by choosing a range of values for the parameters $\alpha$ and $\beta$
referring to the second Ramsey zone. 

\end{twocolumn}

\begin{references}
\bibitem{1.zeng}Shi-Biao Zheng, Phys. Rev. Lett. {\bf 87}, 230404 (2001).
\bibitem{2.Agarwal}G.S. Agarwal, R.R. Puri, and R.P. Singh,
Phys. Rev. A {\bf 56}, 2249 (1997).
\bibitem{3.gsa}G.S. Agarwal, {\em $5^{th}$ Wigner
Symposium}, pp. 313-322, edited by P. Kasperkovitz and D. Grau
(World Scientific, 1998).
\end{references}
\end{document}